\documentclass{article}
\usepackage{spconf}
\usepackage{amssymb}
\usepackage[T1]{fontenc}
\usepackage{placeins}
\usepackage{float}
\usepackage{ragged2e}
\usepackage{fancyhdr}
\usepackage{booktabs}
\usepackage{graphicx}
\usepackage{amssymb,amsmath,bm}
\usepackage{textcomp}
\usepackage{graphicx}
\usepackage{wrapfig}
\usepackage{lipsum}
\usepackage{dblfloatfix}
\usepackage{fixltx2e}
\usepackage{nameref}
\usepackage{amsmath}
\usepackage{graphicx}
\usepackage{varioref}
\usepackage{booktabs}  
\usepackage{siunitx}
\usepackage{numprint}
\usepackage{color}
\usepackage{textcomp}

\usepackage{lipsum}
\begin{document}
	
\name{Rabindra K. Barik\textsuperscript{1}, Harishchandra Dubey\textsuperscript{2}\thanks{\textcolor{blue}{This material is presented to ensure timely dissemination of scholarly and technical work. Copyright and all rights therein are retained by the authors or by the respective copyright holders. The original citation of this paper is:
			 Rabindra K. Barik, Harishchandra Dubey, Kunal Mankodiya, SoA-Fog: Secure Service-Oriented Edge Computing Architecture for Smart Health Big Data Analytics, 5th IEEE Global Conference on Signal and Information Processing GlobalSIP 2017, November 14-16, 2017, Montreal, Canada.}}, Kunal Mankodiya\textsuperscript{3}\thanks{The research discussed in this manuscript was supported by National Institute of Health Grant: R01MH108641.}
\address{\textsuperscript{1} KIIT University, Odisha, India (rabindra.mnnit@gmail.com) \\
\textsuperscript{2} The University of Texas at Dallas, Richardson, TX-75080, USA (harishchandra.dubey@utdallas.edu)\\
\textsuperscript{3}  University of Rhode Island, Kingston, RI-02881, USA (kunalm@uri.edu) \\ 
}
}
\title{SoA-Fog: Secure Service-Oriented Edge Computing Architecture for Smart Health Big Data Analytics}
\maketitle
\begin{abstract}
The smart health paradigms employ Internet-connected wearables for telemonitoring, diagnosis for providing inexpensive healthcare solutions. Fog computing reduces latency and increases throughput by processing data near the body sensor network. In this paper, we proposed a secure service-orientated edge computing architecture that is validated on recently released public dataset. Results and discussions support the applicability of proposed architecture for smart health applications. We proposed~\textit{SoA-Fog} i.e. a three-tier secure framework  for  efficient management of health data using fog devices. It discuss the security aspects in client layer, fog layer and the cloud layer. We design the prototype by using win-win spiral model with use case and sequence diagram. Overlay analysis was performed using proposed framework on malaria vector borne disease positive maps of Maharastra state in India from 2011 to 2014. The mobile clients were taken as test case. We performed comparative analysis between proposed secure fog framework and state-of-the art cloud-based framework.
\end{abstract}
\begin{keywords}
Fog Computing, Service-Oriented Architecture, Smart Health, Big Data Analytics.
\end{keywords}
\vspace{-3mm}
\section{Introduction}
\label{sec:intro}
\vspace{-3mm}
Sharing, storing and processing of public health  data requires secure infrastructure. Health data could be analyzed for locating the area with critical issues of diseases so that proper healthcare facilities could be provided. In many cases, propagation of diseases and ailments are somehow related to geographical location, e.g. Zika Virus in Puerto Rico~\emph{etc.}. Fog computing could be leverage for enhanced analysis of real-world data about diseases and other problems along with the locations~\cite{borthakur2017smart}. Health data are heterogeneous that lead to challenges in integrating it with existing healthcare facilities, interoperability~\emph{etc.} Fog Computing is an emerging solution that provides low-power node for increasing throughput and reducing latency near the edge of various systems at client layer~\cite{bonomi2012fog}. Fog computing requires less cloud storage and transmission power for long-term analysis data. Fog computing has been applied successfully in healthcare and smart cities~\cite{constant2017fog}~\cite{barik2016foggis} ~\cite{dubey2017fog}.  
The present paper has made the following contributions to the secure transmission of health data:
\begin{enumerate}
\item Proposed \textit{SoA-Fog}, a three-tier secure fog computing based framework  that allows communication between client layer, fog layers/nodes and cloud layer for enhanced security features for health data sharing in secure and more efficient way. 
\item Sketched the prototype development by using win-win spiral model. The interaction between the various services modeled by using Unified Modeling Language (UML) with use case and sequence diagrams.
\item Overlay analysis was performed on malaria vector borne disease positive maps of Maharastra state in India from 2011 to 2014 for mobile clients.
\end{enumerate}
\vspace{-3mm} 
\section{Related Works}
\label{sec:background}
\vspace{-3mm} 
\subsection{Fog and Cloud Computing}
\begin{figure*}[!t]
\centering
\includegraphics[width=460pt]{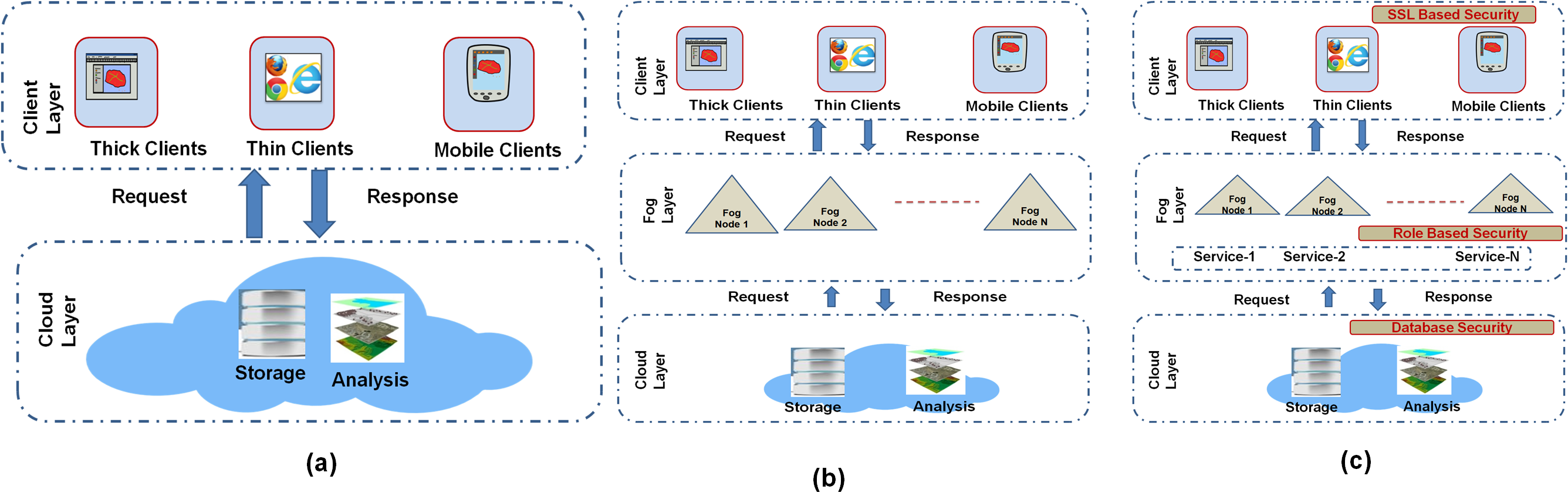}
\caption{(a) Systems' view of Cloud framework with three types of clients (thick, thin and mobile). It has client layer and cloud layer with different web services for analysis and storing of sensitive health data~\cite{r29}; (b)Conceptual diagram of the Fog framework for power-efficient, low latency and high throughput analysis of the geospatial health data; (c) Conceptual overview of \textit{SoA-Fog} for enhanced security features}
\label{fig1}
\end{figure*}

\vspace{-2mm}
Cloud computing has provided ample storage and computational infrastructure for data analysis. It facilitated a transition from desktop to cloud servers. Cloud computing along with other web architectures have created an open environment with shared assets~\cite{r14}. Cloud framework delivered a robust platform in organizations that interrelate tools, technologies and expertise to nurture production, handling and use of geographical data. 
Many cloud platforms uncover the application functionalities through geospatial web services~\cite{r22}. This permit clients to query and update different types of cloud services. It also has provisions of a typical tool to assimilate different cloud applications in the software cloud with enterprise SOA infrastructure.  Figure~\ref{fig1} (a) shows systems' view of Cloud Framework for sharing and storing of health data~\cite{r29}.
In  client-tier layer, there are three types of clients namely mobile, thick and thin. Clients visualize and analyze the geospatial data. Mobile client operates though mobile devices whereas thin clients works on standard web browsers. In thick clients environment, users process or visualize the geospatial data on desktops that requires installation of additional software for full-phase operations~\cite{r31}.The cloud layer is comprised of main geospatial services executed on the servers. It is intermediate between service providers and clients. There are different type of services such as, Web Map Service (WMS), Web Coverage Service (WCS), Web Feature Service (WFS), Web Catalog Service (CSW) and Web Processing Service (WPS) that operates on top of dedicated servers~\cite{r22}.  The detail explanation of every process done by client request, being forward the desire processing service with input of several factors, specifies and provides definite region in leaping box and feedbacks with composite standards.  In cloud layer comprises of storing and analysis of the various geospatial data. System utilizes the layer to store, recover, manipulate and update the geospatial data for long-term analysis. 
%
%
With the technology enhancement of fog computing, it has given the more computing power to the cloud framework. Fog framework has three layers as client tier layer, cloud layer and Fog layer. In client tier, the categories of users have been further divided into thick client, thin client and mobile client environment. Processing of geospatial health data can be possible within these three environments.  Cloud layer is mainly focused on overall storage and analysis of geospatial data. Fog layer works as middle tier between client layer and cloud layer. It has experimentally validated that the Fog layers are characterized by low power consumption, reduced storage requirement and overlay analysis capabilities.In the Fog layer, all the Fog node developed with Intel Edison processor. Fog framework used to assist and hence enhance the capabilities of cloud framework. In Fog framework, fog node processes the data. After processing, it has the ability to send the data to cloud layer for long-term storage and analysis. So, Fog framework enables the more power to the end-users for better performance without computational overhead at cloud layer. Fog framework added privacy benefit where we process the data locally at Fog devices and send only the analysis results to cloud layer. Figure ~\ref{fig1} (b) shows the Conceptual diagram of the fog framework for geospatial health data storage and analysis \cite{barik2016foggis, barik2018fog,barik2017mistgis}. From the above conceptual diagram of cloud and fog framework, it observed that the geospatial data as a key components for data analysis in cloud layer~\cite{r31,r29}. It requires geospatial data from the various components. It led to the concept of geospatial big data that is discussed in the next section.
%
\subsection{Geohealth Big Data}
Big data are data those distribution, diversity, scale and timeliness require the use of new technical architectures and analytics to enable insights that unlock new sources of business value. Big data have included data sets with sizes beyond the ability of commonly used software tools to capture, accurate, manage and process data within an acceptable elapsed time~\cite{r35,lenka2016comparative}. Big data can come in multiple forms. Most of the big data are semi-structured, quasi structured or unstructured, that requires numerous techniques and tools to analyze and process. Analysis of big datasets can discover the new correlations to spot business trends, combat crime and prevent diseases. Big data sets are growing rapidly because they are increasingly gathered by  the information sensing mobile devices, microphones, wireless sensor networks,  cameras,  aerial images and software logs ~\cite{r36}. 
As we know that the reliability, manageability and cost saving are the key important factors in that cloud computing always be one of advantageous over other emerge technology for data processing. But in terms of security and privacy are the main concerns for the processing of sensitive data. Particularly in health sector, data are so sensitivity for further processing and analysis ~\cite{dubey2015fog}. Particularly, for health sector, disease data sharing has been a significant issues for the collaborative preparation, recovery and response stages of numerous disease control mechanism. Disease phenomena are strongly associated with geospatial and related temporal factors. For tackling these situation, Cloud framework has provided dynamic and real-time way to represent disease information through the maps on common browsers ~\cite{bonomi2012fog} ~\cite{barik2016foggis}. So for sharing and analysis of health data in secure way, we have to concentrate with various security issues which has discussed in next section.
\subsection{Security Issues}
With the commencement of cloud computing technology, it has also given so many issues in security and privacy issues. A number of security threats are associated with cloud data services: not only traditional security threats, such as network eavesdropping, illegal invasion, and denial of service attacks, but also specific cloud computing threats, such as side channel attacks, vartualization vulnerabilities, and abuse of cloud services. The following security requirements limit the threats \cite{tang2016ensuring, hafner2009basic}.

Since fog is deemed as a non-trivial extension of cloud, some security and privacy issues in the context of cloud computing, can be foreseen to unavoidably impact fog computing. Security and privacy issues will lag the promotion of fog computing if not well addressed, according to the fact that 74 percent of IT Executives and Chief Information Officers reject cloud in term of the risks in security and privacy . As fog computing is still in its initial stage, there is little work on security and privacy issues \cite{deng2016optimal}. Since fog computing is proposed in the context of Internet of Things (IoT), and originated from cloud computing, security and privacy issues of cloud are inherited in fog computing \cite{perera2017fog}. Client authentication, service security and database security are the prime concern in cloud computing environment. By keeping this on mind, it has been proposed a three tier security framework for sharing of health data across the web \cite{yi2015security, yu2017towards, barik2017foggeo}. From the above related work, it is summarized that, it requires a secure fog computing based framework for sharing and analysis of geospatial big data.	
\section{Prototype Development}
\vspace{-3mm} 
For the prototype development of \textit{SoA-Fog} i.e. Fog-based framework, the primary emphasis is on spiral model of Object Oriented Software Engineering (OOSE) method. In OOSE approach of spiral model, the software development process adopts a sequence of steps including requirements prerequisite plan, analysis, development strategy, operation and testing, complete module and framework observation. The process has basically incremental in nature and each implementation refines the analysis and developing stages through evaluation and testing of a completed module. Further, the incremental development strategy of the proposed framework which allows the problem of constructing this framework to be tackled in smaller, more controllable portions of increasing complexity. So there are four phases in that to be defined in \textit{SoA-Fog}. Phase I deals with the proposed model of \textit{SoA-Fog} framework. Phase II describes about the use case and sequence diagram for proposed \textit{SoA-Fog}. Phase III and Phase IV explain the overlay analysis of the geospatial data on mobile client environment in \textit{SoA-Fog} framework and comparison analysis for Cloud framework and \textit{SoA-Fog} framework. 
\begin{figure*}[!t]
\centering
\includegraphics[width=460pt]{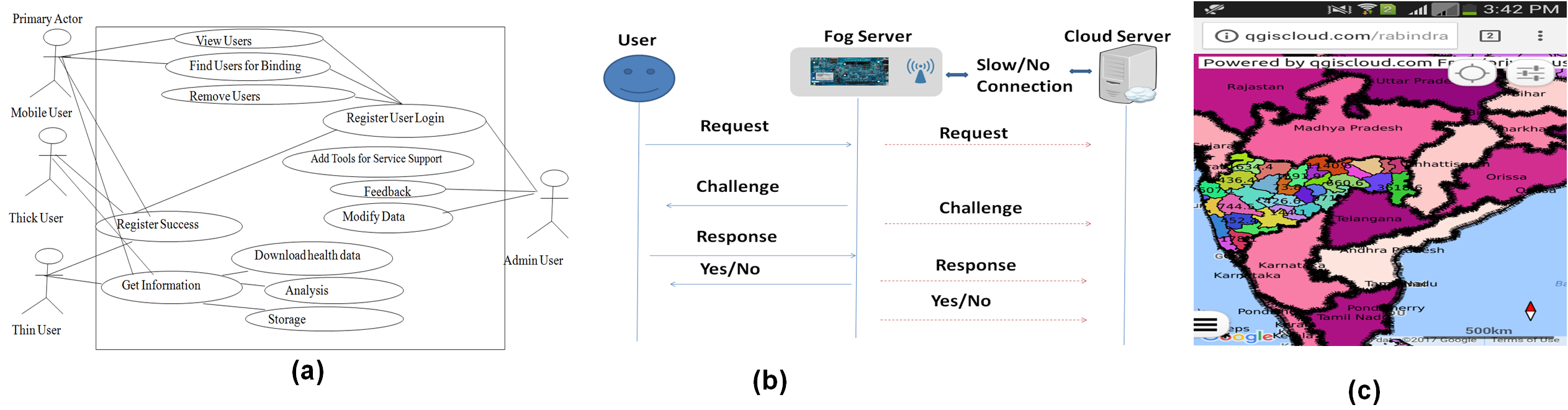}
\caption{(a) Conceptual diagram of the Fog framework for power-efficient, low latency and high throughput analysis of the geospatial health data; (b) Conceptual overview of \textit{SoA-Fog} for enhanced security features (c) Overlay operation on mobile client environment in qgis cloud~\cite{r59} }
\label{fig2}
\end{figure*}	
\vspace{-4mm}
\subsection{Proposed Framework}
\vspace{-3mm} 
This section describes various components of the proposed \textit{SoA-Fog} framework and discusses the methods implemented in it.The main focus on \textit{SoA-Fog} has been  use of a practical approach to explore and extend the concept of security approaches for Fog computing in health sector. It should provide an effective and efficient means of sharing health related data on the web. Figure ~\ref{fig1} (c) shows the proposed three- tier secure service oriented Fog computing framework of health data resources in which the basic over view of service provider, service consumer and catalog service are being shown.

In the \textit{SoA-Fog} framework, it is proposed to achieve the principle of CIA. Confidentiality can be achieved by SSL based security integration. Role based security is meant to focus on integrity of services whereas database security is to focus on the availability of the data to the authenticated user. In the proposed \textit{SoA-Fog} framework, it is proposed 3-tier security mechanism, the middle layer i.e. fog layer is technically meant to be role base access control mechanism. The easy to implement mechanisms like discretionary access control mechanism and mandatory access control mechanism can be used. Role base access control mechanism in a preventive way accesses the data tier and ultimately the data will reach the application layer passing through the security mechanism of the data tier and cloud layer. The user role is defined at the very beginning at the client layer by providing authorized access after authenticated verification of the user identity.
		
In addition, it is expected that each phase would reveal a unique features related to the requirements of infrastructure and enable exploration of the interfaces between fog framework components. The requirements stage of application design aims to specify the behavior of the framework from perspective of a user. From the above defined \textit{SoA-Fog} framework, it is described the details of use case model and sequence diagram. Figure ~\ref{fig2} (b) and ~\ref{fig2} (c) have shown the use case and sequence digram of \textit{SoA-Fog} framework. In the proposed \textit{SoA-Fog} framework, it is more secure for sharing of health data than cloud based framework. So the next result and discussions section describes about overlay analysis and the comparison analysis of existing cloud framework with \textit{SoA-Fog} framework by taking suitable parameters. 
\begin{table}[!t]
\begin{center}
\caption{Comparison of cloud framework and \textit{SoA-Fog} Framework} \label{tab:table3} 
\vspace{-3mm}
\begin{tabular}{|p{2cm}|p{3cm}|p{2.8cm}|}
			\hline
			\textbf{Features} & \textbf{Cloud Framework} & \textbf{\textit{SoA-Fog}} \\ \hline
			Bandwidth Requirements and  Internet Connectivity  & Requires clients to have network connectivity to the cloud server and bandwidth requirements grow with the total amount of health data generated by clients.
			& Operates autonomously to provide uninterrupted services even no or intermittent Internet connectivity and network bandwidth requirements grow with total the amount of data that need to be process and sent to the cloud server after being authenticated by the fog layer. \\ \hline
			\textbf{Size }
			& Processing has done with large amount of health data at a time and each typically contains tens of thousands of integrated servers & Fog node in each location can be small and work with role based access mechanism. \\\hline
			\textbf{Server Locations} & Requires centralized server in a small number of big data centers distributed environment
			& Requires distributed servers in many locations and over large geographical areas, closer to users along with fog-to-cloud range\\
			\hline
		\end{tabular}
	\end{center}
\end{table}
\vspace{-4mm}
\section{Results~\& Discussions}
\vspace{-3mm}
In this section, data analysis particularly overlay analysis is performed for malaria vector borne disease positive maps of Maharastra, India. It has been found that 2 number of shape files related to malaria information mapping are overlaying with google satellite layer. In the present study, it has been used the malaria death mapping data of Maharastra from 2011-2014; has been processing in \textit{SoA-Fog}.The overlay analysis of various vector data and raster data of particular area has been performed. Initially, the developed datasets have been opened with Quantum GIS; desktop based GIS analysis tools, and performed some join operations in mobile client environment. In Quantum GIS, plugin named as QGISCloud has installed. The said plugin has the capability of storing various raster and vector data set in cloud database for further overlay analysis. After storing in cloud database, it generates the mobile and thin client link for visualization of both vector and raster data set. Figure~\ref{fig2} (c) shows the overlay analysis on mobile client environment. It observes that the overlay analysis is a useful technique for visualization of health data.
%
Both Cloud and \textit{SoA-Fog} frameworks have specific meaning for a service range with in the cloud computing and client tiers which provide the mutual benefit to each other and interdependent services that leads to the greater storage capacity, control and communication with in the specified range. Table ~\ref{tab:table3}  outlines the comparison characteristics of cloud and \textit{SoA-Fog} framework.
%
\vspace{-5mm}
\section{Conclusions}
\label{sec:Conclusions}
\vspace{-3mm}
In this study, we proposed \textit{SoA-Fog} framework for enhanced analysis of geo-health data. Intel Edison was used as Fog computer in developed prototypes. Fog devices reduced the storage requirements, transmission power leading to overall efficiency. We performed a case study using Geo-health data of malaria vector borne disease positive maps of Maharastra state in India.  We performed the overlay analysis using proposed architecture. In this way, the fog devices add edge intelligence in geo-health data analysis by introducing local processing within cloud-based computing environments.
%
%
\bibliographystyle{IEEEbib}
\bibliography{refs2}

\begin{thebibliography}{10}

\bibitem{borthakur2017smart}
D.~Borthakur, H.~Dubey, N.~Constant, L.~Mahler, and K.~Mankodiya,
\newblock ``Smart fog: Fog computing framework for unsupervised clustering
  analytics in wearable internet of things,''
\newblock in {\em 5th IEEE Global Conference on Signal and Information
  Processing}, 2017.

\bibitem{bonomi2012fog}
Flavio Bonomi, Rodolfo Milito, Jiang Zhu, and Sateesh Addepalli,
\newblock ``Fog computing and its role in the internet of things,''
\newblock in {\em ACM MCC}, 2012.

\bibitem{constant2017fog}
Nicholas Constant~etal.,
\newblock ``{Fog-Assisted wIoT: A Smart Fog Gateway for End-to-End Analytics in
  Wearable Internet of Things},''
\newblock in {\em IEEE HPCA 2017}.

\bibitem{barik2016foggis}
R.~K. Barik, H.~Dubey, A.~B. Samaddar, R.~D. Gupta, and P.~K. Ray,
\newblock ``Foggis: Fog computing for geospatial big data analytics,''
\newblock {\em arXiv preprint arXiv:1701.02601}, 2016.

\bibitem{dubey2017fog}
H.~Dubey, N.~Constant, A.~Monteiro, M.~Abtahi, D.~Borthakur, L.~Mahler, Y.~Sun,
  Q.~Yang, and K.~Mankodiya,
\newblock ``Fog computing in medical internet-of-things: Architecture,
  implementation, and applications,''
\newblock in {\em Handbook of Large-Scale Distributed Computing in Smart
  Healthcare}. 2017, Springer International Publishing AG.

\bibitem{r14}
C.~Yang, Q.~Huang, Z.~Li, K.~Liu, and F.~Hu,
\newblock ``Big data and cloud computing: innovation opportunities and
  challenges,''
\newblock {\em International Journal of Digital Earth}, vol. 10, no. 1, pp.
  13--53, 2017.

\bibitem{r22}
R.~A.~AL Kharouf, A.~R. Alzoubaidi, and M.~Jweihan,
\newblock ``An integrated architectural framework for geoprocessing in cloud
  environment,''
\newblock {\em Spatial Information Research}, pp. 1--9, 2017.

\bibitem{r29}
K.~Evangelidis, K.~Ntouros, S.~Makridis, and C.~Papatheodorou,
\newblock ``Geospatial services in the cloud,''
\newblock {\em Computers \& Geosciences}, vol. 63, pp. 116--122, 2014.

\bibitem{r31}
P.~Yue, H.~Zhou, J.~Gong, and L.~Hu,
\newblock ``Geoprocessing in cloud computing platforms--a comparative
  analysis,''
\newblock {\em International Journal of Digital Earth}, vol. 6, no. 4, pp.
  404--425, 2013.

\bibitem{barik2018fog}
R.~K. Barik, H.~Dubey, C.~Misra, D.~Borthakur, N.~Constant, S.~A. Sasane, R.~K.
  Lenka, B.~S.~P. Mishra, H.~Das, and K.~Mankodiya,
\newblock ``Fog assisted cloud computing in era of big data and
  internet-of-things: Systems, architectures and applications,''
\newblock in {\em Cloud Computing for Optimization: Foundations, Applications,
  Challenges}, p.~23. Springer, 2018.

\bibitem{barik2017mistgis}
R.~Barik, H.~Dubey, R.~K. Lenka, K.~Mankodiya, T.~Pratik, and S.~Sharma,
\newblock ``Mistgis: Optimizing geospatial data analysis using mist
  computing,''
\newblock in {\em International Conference on Computing Analytics and
  Networking (ICCAN 2017)}. Springer, 2017.

\bibitem{r35}
J.~Andreu-Perez, C.~C.~Y. Poon, R.~D. Merrifield, S.~T.~C. Wong, and G.~Yang,
\newblock ``Big data for health,''
\newblock {\em IEEE journal of biomedical and health informatics}, vol. 19, no.
  4, pp. 1193--1208, 2015.

\bibitem{lenka2016comparative}
R.~K. Lenka, R.~K. Barik, N.~Gupta, S.~M. Ali, A.~Rath, and H.~Dubey,
\newblock ``Comparative analysis of spatialhadoop and geospark for geospatial
  big data analytics,''
\newblock in {\em 2nd International Conference on Contemporary Computing and
  Informatics (IC3I 2016)}. IEEE, 2016.

\bibitem{r36}
J.~Lee and M.~Kang,
\newblock ``Geospatial big data: challenges and opportunities,''
\newblock {\em Big Data Research}, vol. 2, no. 2, pp. 74--81, 2015.

\bibitem{dubey2015fog}
H.~Dubey, J.~Yang, N.~Constant, A.~M. Amiri, Q.~Yang, and K.~Makodiya,
\newblock ``Fog data: Enhancing telehealth big data through fog computing,''
\newblock in {\em Fifth ASE BigData 2015, Kaohsiung, Taiwan}. ACM.

\bibitem{tang2016ensuring}
J.~Tang, Y.~Cui, Q.~Li, K.~Ren, J.~Liu, and R.~Buyya,
\newblock ``Ensuring security and privacy preservation for cloud data
  services,''
\newblock {\em ACM Computing Surveys (CSUR)}, vol. 49, no. 1, pp. 13, 2016.

\bibitem{hafner2009basic}
M.~Hafner and R.~Breu,
\newblock ``Basic concepts of soa security,''
\newblock {\em Security Engineering for Service-Oriented Architectures}, pp.
  27--45, 2009.

\bibitem{deng2016optimal}
R.~Deng, R.~Lu, C.~Lai, T.~H. Luan, and H.~Liang,
\newblock ``Optimal workload allocation in fog-cloud computing towards balanced
  delay and power consumption,''
\newblock {\em IEEE Internet of Things Journal}, 2016.

\bibitem{perera2017fog}
C.~Perera, Y.~Qin, J.~C. Estrella, S.~Reiff-Marganiec, and A.~V. Vasilakos,
\newblock ``Fog computing for sustainable smart cities: A survey,''
\newblock {\em arXiv preprint arXiv:1703.07079}, 2017.

\bibitem{yi2015security}
S.~Yi, Z.~Qin, and Q.~Li,
\newblock ``Security and privacy issues of fog computing: A survey,''
\newblock in {\em International Conference on Wireless Algorithms, Systems, and
  Applications}. Springer, 2015, pp. 685--695.

\bibitem{yu2017towards}
Z.~Yu, M.~H. Au, Q.~Xu, R.~Yang, and J.~Han,
\newblock ``Towards leakage-resilient fine-grained access control in fog
  computing,''
\newblock {\em Future Generation Computer Systems}, 2017.

\bibitem{barik2017foggeo}
R.~K. Barik, H.~Dubey, R.~K. Lenka, N.V.R. Simha, S.~A. Sasane, C.~Misra, and
  K.~Mankodiya,
\newblock ``Fog computing-based enhanced geohealth big data analysis,''
\newblock in {\em 2017 International Conference on Intelligent Computing and
  Control (I2C2)}. IEEE, 2017.

\bibitem{r59}
``Available on :
  http://qgiscloud.com/rabindrabarik2016/\\malaria?mobile=true,''
\newblock Accessed on: 13th February 2017.

\end{thebibliography}
\end{document}